\documentclass[twocolumn]{revtex4}

\usepackage{graphicx}
\usepackage{dcolumn}
\usepackage{bm}

\usepackage{amssymb}

\begin{document}

\title{Quantum mechanics and geodesic deviation in the brane world}

\author{S. M. M. Rasouli}
\author{A. F. Bahrehbakhsh}
\author{S. Jalalzadeh}
\email{s-jalalzadeh@sbu.ac.ir}
\author{M. Farhoudi}
\affiliation{Department of Physics, Shahid Beheshti University, G. C., Evin, Tehran 19839, Iran
}%

\date{\today}

\begin{abstract}
We investigate the induced geodesic deviation equations in the brane world models, in
which all the matter forces except gravity are confined on the 3-brane. Also, the Newtonian limit of
induced geodesic deviation equation is studied. We show that in the first Randall-Sundrum model
the Bohr-Sommerfeld quantization rule is as a result of consistency between the geodesic and
geodesic deviation equations. This indicates that the path of test particle is made up of integral
multiples of a fundamental Compton-type unit of length $h/mc$.
\vspace{5mm}\noindent\\
\end{abstract}

\maketitle

\section{}
The mission to formulate a consistent quantum theory
of gravity has maintained physicists busy since the first
attempt by Rosenfeld in 1930. In spite of much work,
no definitive progress has been made. Nowadays, there
are many interesting attempts to quantize gravity. In this
paper we take an opposite direction: we will show that
quantum objects can be constructed from gravitationalgeometrical
effects. Actually, the idea of geometrization
of quantum mechanics has been considered in different
approaches. For example, one can increase the number of
dimensions of spacetime in Kaluza-Klein (KK) models of
gravity~\cite{1}, Weylian spacetime~\cite{2}, scalar-tensor theories of
gravity~\cite{3} or other possible extensions of Einstein general
relativity. Recently, it has been shown that the existence of
non-compact extra dimensions leads to quantum effects in
the classically induced 4-dimensional (4D) physical entities~\cite{4}. In~\cite{5}, to construct semi-classical quantum gravity
from geometric properties of brane, the authors have used
the Induced Matter Theory (IMT) which is an extension
of the KK theory. In this approach, not only the gauge
fields are unified with gravity (geometry) but also the
matter fields are unified with gravity and have geometrical
origin, constructed from extrinsic curvature~\cite{6}. The origin
of quantum effects in fact is the fluctuation of matter fields
around 4D spacetime.

In this paper we discuss the existence of quantum effects
in the most famous model of brane gravity. In this model and its extensions, the presence of non-compact extra
dimension is not in fact for the unification shame, but
for the explanation of hierarchy problem without using
supersymmetry~\cite{RStwo99}.

The idea that our familiar 4D spacetime is a hypersurface
(brane) embedded in a 5D bulk has been experiencing
a phenomenal interest during the last decade. The
behavior of geodesics and the Newtonian limit of linearized
gravity for the Randall-Sundrum (RS) and an alternative
brane background have been investigated extensively~\cite{MVV00}.
Also,  Ref.~\cite{Y00} has  looked into the geodesic motions of
a test particle in the bulk spacetime
in RS scenario.
The induced
4D geodesic equation on the brane, to which we assume
that the matter fields except gravity is confined, is given
by~\cite{JS05}
\begin{eqnarray}\label{geodesic.eq}
\frac{d^2x^\mu}{d\tau^2}+\Gamma^\mu_{\alpha\beta}
\frac{dx^\alpha}{d\tau}\frac{dx^\beta}{d\tau}=0\,,
\end{eqnarray}
where $\tau$ is the proper time defined on the brane and $\Gamma^\mu_{\alpha\beta}$
are 4D Christoffel symbols derived from the induced metric. (Here and throughout we shall use $A, B = 0,1,2,3,5$ to denote 5D coordinates,
$\mu,\nu = 0,1,2,3$ to denote the standard 4D ones and $\bar
A=1,2,..,5, \ \bar{\mu}=1,2,3$ denotes spacelike counterparts).

Note that in Eq.~(\ref{geodesic.eq}) 
 the effect of the existence of bulk space is hidden in the
induced metric which one can obtain  via induced Einstein field
equations~\cite{SMS00}.
To obtain the induced geodesic (\ref{geodesic.eq}), we usually
start from the geodesic equation of a test particle in the
bulk space and then reduce it to the 4D hypersurface.
One can use the same procedure to acquire
induced geodesic deviation (GD) equation. For example in Kaluza-Klein
theory authors of  \cite{KMMH00} used the same method to obtain GD on this
kind of compact models.
Hence, we start with the GD equation associated to the bulk space, namely
\begin{eqnarray}\label{GD.eq}
\frac{^{(5)}D^2\xi^A}{DS^2}=
{\cal R}^A_{\,\,\, BCD}\frac{dx^B}{dS}\frac{dx^C}{dS}\xi^D,
\end{eqnarray}
where ${\cal R}^A_{\,\,\, BCD}$ is the Reimann tensor for the bulk space, $\xi^A$
is an infinitesimal GD vector, $D/DS$ denotes the pull--back
of covariant derivatives and $S$ is an affine parameter for the bulk space.
To induce  Eq.~(\ref{GD.eq}) on the brane we need induced
components of the Reimann tensor of the bulk space on the brane, \emph{i.e}. Gauss-Codazzi
equations.
In the Gaussian normal frame, explicit calculation directly gives
\begin{eqnarray}\label{Reimann1}
{\cal R}^\mu_{\,\,\, \alpha\beta\gamma}=
R^\mu_{\,\,\, \alpha\beta\gamma}
+K_{\alpha\beta}K^\mu_{\,\,\, \gamma}-K_{\alpha\gamma}K^\mu_{\,\,\,
\beta},
\end{eqnarray}
and
\begin{eqnarray}\label{Reimann2}
{\cal R}^\mu_{\,\,\, 4\alpha 4}=K^\mu_{\,\,\, \alpha,4}-K^\sigma_{\,\,\, \alpha}K_\sigma^{\,\,\,
\mu},
\end{eqnarray}
where $R^\mu_{\,\,\, \alpha\beta\gamma}$ is 4D Reimann tensor
and $K_{\mu\nu}$ denotes
the extrinsic curvature.
Inserting Eqs.~(\ref{Reimann1})
and (\ref{Reimann2}) into the Eq.~(\ref{GD.eq}) gives
\begin{eqnarray}\label{Hyp.GD}
\begin{array}{cc}
\frac{D^2\xi^\mu}{DS^2} =\\
\\
\left (R^\mu_{\,\,\, \alpha\beta\gamma}+ K_{\alpha\beta}K^\mu_{\,\,\,
\gamma} - K_{\alpha\gamma}K^\mu_{\,\,\, \beta}\right)
\frac{dx^\alpha}{dS}\frac{dx^\beta}{dS}\xi^\gamma+\\
\\
\epsilon \left(K^\mu_{\,\,\, \alpha,4}-K_{\alpha\sigma}
K^{\sigma\mu} \right)\left[\frac{dx^4}{dS}\frac{dx^\alpha}{dS}
\xi^4 - (\frac{dx^4}{dS})^2\xi^\alpha \right].
\end{array}
\end{eqnarray}
 Now the derivatives with respect
to the 5D line element $dS$, should be replaced by the
derivatives with respect to the 4D Affine parameter.
To attend to this aim, we rewrite Eq.~(\ref{Hyp.GD}) with a general parameter
$\lambda$, which parameterizes 4D motion as
\begin{eqnarray}\label{rel4,5}
\frac{^{(5)}D^2\xi^\mu}{DS^2}=
\left(\frac{d\lambda}{dS}\right)^2
\frac{^{(5)}D^2\xi^\mu}{D\lambda^2}
+ \frac{d\lambda}{dS}\frac{d}{d\lambda}
(\frac{d\lambda}{dS})\frac{^{(5)}D\xi^\mu}{D\lambda}\,,
\end{eqnarray}
where the relation between $5$ and $4$--dimensional
covariant differentiations
is given by
\begin{eqnarray}\label{simplifyHGD1}
\begin{array}{cc}
\frac{^{(5)}D\xi^\mu}{D\lambda}=
 \frac{d\xi^\mu}{d\lambda}
 +^{(5)}\Gamma^\mu_{AB}\frac{dx^A}{d\lambda}\xi^B
=\frac{D\xi^\mu}{D\lambda}\\
\\
-\epsilon K^\mu_{\,\,\, \alpha}
\frac{dx^\alpha}{d\lambda}\xi^4-\epsilon K^\mu_{\,\,\,
\alpha}\frac{dx^4}{d\lambda}\xi^\alpha,
\end{array}
\end{eqnarray}
so that in the second equality, 5D Christoffel symbols have been replaced
by the 4D counterparts using their relations obtained in Ref.~\cite{JS05}.
Now, from Eqs.~(\ref{Hyp.GD}), (\ref{rel4,5}) and (\ref{simplifyHGD1}) we obtain
\begin{eqnarray}\label{simplifyHGD2}
\begin{array}{cc}
\frac{D^2\xi^\mu}{D\lambda^2}=R^\mu_{\,\,\, \alpha\beta\gamma}
\frac{dx^\alpha}{d\lambda}\frac{dx^\beta}{d\lambda}\xi^{\gamma}+\\
\\
\left(K_{\alpha\beta}K^\mu_{\,\,\,
\beta} -K_{\alpha\gamma}K^\mu_{\,\,\, \beta}\right)
\frac{dx^\alpha}{d\lambda}\frac{dx^\beta}{d\lambda}\xi^\gamma
+\\
\\
\epsilon \left(K^\mu_{\,\,\, \alpha,4} -
K^{\rho}_{\,\,\, \alpha}K^\mu_{\,\,\,
\rho}\right)\left[\frac{dx^\alpha}{d\lambda}
\frac{dx^4}{d\lambda}\xi^4-(\frac{dx^4}{d\lambda})^2\xi^\alpha
\right]-\\
\\
\left[\frac{D\xi^\mu}{D\lambda}-\epsilon K^\mu_{\,\,\, \alpha}
\left(\frac{dx^\alpha}{d\lambda}\xi^4+\frac{dx^4}
{d\lambda}\xi^\alpha\right)\right]\left(\frac{d\lambda}
{dS}\right)^{-1}\frac{d}{d\lambda}(\frac{d\lambda}{dS}).
\end{array}
\end{eqnarray}
The above induced GD equation
can be used in  various brane models.
For example, in the Induced Matter Theory (IMT) \cite{4,WP92},
the test particles are not in general,
confined to the specific fixed brane \cite{J07}.
In this case, since the extra component of velocity
of the test particle, $u^4=dx^4/d\lambda$,
does not vanish,  all the extra terms
in the right hand side of Eq.~(\ref{simplifyHGD2}) will be present.
Another important point in the IMT is choice of
$\lambda$, the parameterization of
the path. Usually, in the literature
has been assumed that the line element of the
brane, which is defined here as the proper
time ``$d\tau$'', is logical and convenient.
However, the non-integrability
property of induced physical quantities on the brane dictates
that the parameterization of the 
path is, in general, deferent from the 4D proper time \cite{J07}.
On the other hand, in the brane phenomenological models where matter field
has been confined on the
fixed brane, the 4D proper time defined on the
brane is required as a suitable parameterization of motion.
In this paper, we would like to study  GD in the
brane models based on the Horava and Witten
theory \cite{HW96}, hence, we will assume that all the matter fields,
except gravity, are confined on
the fixed brane. Therefore, in Eq.~(\ref{simplifyHGD2}), $d\lambda$ will be
substituted by $d\tau$, the proper time defined
on the brane. Furthermore, we  assume that the velocity of test particles
along the extra dimension vanishes.
Imposing the above assumptions on Eq.~(\ref{simplifyHGD2}), we obtain
\begin{eqnarray}\label{1-14}
\begin{array}{cc}
\frac{D^2\xi^\mu}{D\tau^2} = R^\mu_{\,\,\, \alpha\beta\gamma}u^{\alpha} u^\beta
\xi^\gamma +\\
\\
 \left(K_{\alpha\beta}K^\mu_{\,\,\, \gamma}-K_{\alpha\gamma}K^\mu_{\,\,\,
\beta}\right)u^\alpha u^\beta \xi^\gamma,
\end{array}
\end{eqnarray}
where $u^\alpha=\frac{dx^\alpha}{d\tau}$ denotes
$4$-velocity of the test particles defined
on the brane. 

In general relativity,
the Newtonian limit of GD equation leads
us to the form of the field Equations ~\cite{MVV00}.
Hence we derive and analyze the
Newtonian limit of Eq.~(\ref{simplifyHGD2}).
We elaborate tensor equation~(\ref{GD.eq}) in the local rest
frame for one of the two test particles $A_1$ and $A_2$
with coordinates $x^A(s,\eta)$ and $x^A(s,\eta+\delta\eta)$, respectively.
In this frame ${\cal G}_{AB} = \eta_{AB}$ and $dS=dt$. This means that $A_1$
promotes its clock to the master clock indicating coordinates time. Also,
${^{(5)}D}/{DS} =d/dt$, $x^A=(t,0,0,0,0)$ and $u^A = (1,0,..,0)$. We are left with
\begin{eqnarray}\label{Newton1}
 \frac{d^2\xi^{\bar{A}}}{dt^2}={\cal R}^{\bar{A}}_{\,\,\, 00\bar{B}}\xi^{\bar{B}}
 \hspace{1cm} (\bar{A}=1,2,3,4).
 \end{eqnarray}
At this point $A_1$ recalls that according
to the classical mechanics both he
and $A_2$ move in a stationary gravitational
field: ${\bf{\ddot{r}}}_{A_1}={\bf{F}}({\bf{r}}_{A_1})$
and  ${\bf{\ddot{r}}}_{A_2}={\bf{F}}({\bf{r}}_{A_2})$.
Setting $\xi^B = {\bf{r}}^B_{A_2}-{\bf{r}}^B_{A_1}$ gives
\begin{eqnarray}\label{Newton2}
\begin{array}{cc}
\frac{d^2\xi^{\bar A}}{dt^2}=F^{\bar A}({\bf{r}}_A+{\xi})-F^{\bar A}({\bf{r}}_A)\simeq \\
\\
F^{\bar A}_{\,\,\,
,\bar B}\xi^{\bar B}=-\Phi^{,{\bar A}}_{\,\,\,,{\bar B}}\xi^{\bar B},
\end{array}
\end{eqnarray}
where $\Phi$ is the gravitational potential in the bulk space.
Comparing Eqs.~(\ref{Newton1}) and  (\ref{Newton2}) gives
\begin{eqnarray}\label{1-17}
{\cal R}^{\bar{A}}_{\,\,\, 00\bar{B}}=-\Phi^{,\bar{A}}_{\,\,\, ,\bar{B}}.
\end{eqnarray}
Now, using this equation and recalling
Eqs.~(\ref{Reimann1}) and (\ref{Reimann2})
we find
\begin{eqnarray}\label{New-Reimann}
\begin{array}{cc}
R^{\bar{\mu}}_{\,\,\, 00\bar{\mu}}+K_{00}K-K_{0\bar{\mu}}K^{\bar{\mu}}_{\,\,\,
0}-K_{,5}-\\
\\
K^{\bar{\mu}}_{\,\,\, \bar{\nu}}K_{\bar{\mu}}^{\,\,\, \bar{\nu}}=-\Phi^{,\bar{A}}_{\,\,\,
,\bar{A}} \hspace{0.5cm}(\bar{\mu}=1,2,3).
\end{array}
\end{eqnarray}
The classical field equations in the bulk space is
\begin{eqnarray}\label{1-19}
\Phi_{,\bar{A}\bar{A}}=-\Lambda+(-\sigma+k^2_5\rho)\delta(x^5)\,,
\end{eqnarray}
where according to the spirit of brane models,
we have assumed existence of the bulk
cosmological constant $\Lambda$,
tension of brane $\sigma$ and the matter density $\rho$.
Consequently, we obtain
\begin{eqnarray}\label{1-20}
\begin{array}{cc}
R^{\bar{\mu}}_{\,\,\, 00\bar{\mu}}+K_{00}K-K_{0\bar{\mu}}K^{\bar{\mu}}_{\,\,\,
0}-K_{,5}-K^{\bar{\mu}}_{\,\,\, \bar{\nu}}K_{\bar{\mu}}^{\,\,\, \bar{\nu}}=\\
\\
-\Lambda+(-\sigma+k^2_5\rho)\delta(x^4).
\end{array}
\end{eqnarray}
Integration along normal direction gives
the Newtonian limit of the Israel junction condition as
\begin{eqnarray}\label{1-21}
\left[K\right] = -k^2_5\rho+\sigma\,,
\end{eqnarray}
where $[X]:=\lim_{x^4\rightarrow 0^{+}}X-\lim_{x^4\rightarrow 0^{-}}X$. Also,
if we impose the $Z_2$ symmetry then we obtain
\begin{eqnarray}\label{1-22}
K^+=\frac{1}{2}(k^2_5\rho-\sigma),
\end{eqnarray}
which is the Newtonian version
of Israel junction condition obtained
in \cite{HW96}. Now, we obtain the GD
equation in the RS brane world
scenario. In the RS scenario,
it has been proposed a 5D bulk space,
which is described by the metric \cite{RStwo99,RSone99}
\begin{eqnarray}\label{RS-metric}
dS^{2}=e^{-2k|y|}\eta_{\mu\nu}dx^{\mu}dx^{\nu}+dy^{2}\,,
\end{eqnarray}
where $y=r\phi$ signifies the extra
spacelike dimension with compactification
radius $r$, $k=\sqrt{-\Lambda/12M^3}$
and $\Lambda$ is the bulk cosmological constant, and $M$ is 
fundamental 5D Plank scale. The factor $e^{-2k\left\vert y \right\vert}$ is
called warp factor and the
geometry of the extra dimension
is orbifolded by $S^1/Z_2$.
In the RSI scenario it can be shown that even
if Higgs or any other mass
parameter in the 5D Lagrangian is of the order
of Planck scale, $m_0\simeq 10^{16}$ TeV,
on the visible brane, it gets warped
by a factor of the form
\begin{eqnarray}\label{M.eq}
m=m_0e^{-kr\pi}.
\end{eqnarray}
Thus by assuming $kr=11.84$, one gets $m\simeq 1$ TeV.
Using RSI metric (\ref{RS-metric}) we obtain
\begin{eqnarray}\label{Sim.RS}
K_{\mu\nu}=k\frac{|y|}{y}e^{-2k|y|}\eta_{\mu\nu}.
\end{eqnarray}
The constant slices at $y=0$ and $y=r\pi$ are
known as the hidden and visible
branes respectively, which the observable universe being
identified with latter. Therefore, the
GD equation (\ref{GD.eq}) on the visible brane becomes
 \begin{eqnarray}\label{1-26}
\frac{D^{2}\xi^{\mu}}{D\tau^{2}}=
\ddot{\xi}^{\mu}=k^2e^{-2\pi kr}(\eta_{\alpha\beta}\eta^{\mu}_{\
\gamma}-\eta_{\alpha\gamma}\eta^{\mu}_{\
\beta})u^{\alpha}u^{\beta}\xi^{\gamma},
\end{eqnarray}
where a dot denotes derivative with
respect to the brane proper time.
On the other hand, solving the geodesic
equation (\ref{geodesic.eq})
on this brane model gives the constant 4-velocity
of test particle as $u^{\mu}=const\,,$
which shows that the initially parallel  geodesics
will always remain parallel  as a property of $4D$ Minkowski
spacetime. The solution of equation (\ref{1-26}) for massive test particles
is
\begin{eqnarray}\label{1-30}
\xi^\mu=f^\mu e^{ike^{-\pi kr}\tau},
\end{eqnarray}
where $f^\mu$ is the integration constant.
Equation (\ref{1-30}) implies that the distance between two
geodesics oscillate contrary to the  geodesic equation. The consistency of this solution with geodesic equation
then impose the following restriction \begin{eqnarray}\label{1-31}
cke^{-\pi kr}\tau=n\pi , \ n=0,1,2,...,
\end{eqnarray}
where $c$ is the speed of light which
 is not considered, here, to be unity.
Also it is well known
that
\begin{eqnarray}\label{1-32}
\int p_{\mu}dx^{\mu}=\int mu_{\mu}dx^{\mu}=
\int m\left(\frac{ds}
{d\tau}\right)^2d\tau=mc^2\tau,
\end{eqnarray}
where $p_\mu$ is the induced  4-momentum of the test particle and $m$ is
the rest mass. Comparing Eqs. (\ref{1-31}) and (\ref{1-32}) gives 
\begin{eqnarray}\label{1-33}
\int p_{\mu}dx^\mu=n\pi\frac{mce^{\pi kr}}{k}.
\end{eqnarray}
Replacing $m$ from Eq.~(\ref{M.eq})
into the above equation, and 
if we set $k\sim 1/l_{Pl}$, Eq.~(\ref{1-33}) reduces to
\begin{eqnarray}\label{1-37}
\int p_{\mu}dx^\mu=nh.
\end{eqnarray}
Which is  similar to the old quantum theory quantization condition but
is less stringent, for the old quantum conditions were the integration being taken for a closed curve. On the other
hand, Eq. (\ref{1-31}) leads to
\begin{eqnarray}\label{1-38}
\tau=n\frac{h}{mc^2},
\end{eqnarray}
indicating that the proper time of the test particle is made up of integral multiples of a fundamental unit of length $h/mc^2$. This result  suggests is that the world-line of the test particle is to be considered as made up of these units of length, nothing less being observable directly or indirectly in experiment. Note that according to the \cite{flint} it could be concluded from (\ref{1-38})
that the smallest interval of time and distance then are given by
\begin{eqnarray}\label{1-39}
\begin{array}{cc}
\delta t=\frac{h}{mc^2}\frac{1}{\sqrt{1-\beta^2}},\\
\\
\delta l=\frac{h}{mc}\frac{\beta}{\sqrt{1-\beta^2}},
\end{array}
\end{eqnarray}
where $\beta=\frac{v}{c}$ and the following uncertainty relations
\begin{eqnarray}\label{1-40}
\Delta p_\mu\Delta x_\mu \sim \frac{2h}{n-1}.
\end{eqnarray}
 Note that in relations (\ref{1-39}) both of them are depend upon the velocity
 of the test particle. For velocities approaching the velocity of light they
 become  very large which means that it is impossible to measure intervals
 of time and length in association with such rapidly moving particles. Hence
 it seems that the deduction from the existence of a least proper time is
 that any accurate measurements on a particle moving with such velocity would
 be impossible. Also in equation (\ref{1-40}) the worst case is for $n=1$,
 but this is no practical significance   
for it corresponds to an observation of one fundamental unit of length which
is recorded as corresponding to zero proper time. In this uncertainty relation
for a  large amount of $n$, the right hand side of (\ref{1-40}) vanish, i.e. this equation naturally contains classical limit. Since the minimum length and time intervals that can be measured are given by (\ref{1-39}) then the maximum uncertainly on 3-momentum and energy becomes
\begin{eqnarray}\label{1-41}
\begin{array}{cc}
\delta p\sim \frac{2mc}{n-1}\frac{\sqrt{1-\beta^2}}{\beta},\\
\\
\delta E\sim \frac{2mc^2}{n-1}{\sqrt{1-\beta^2}}.
\end{array}
\end{eqnarray}
The conclusion is that the above uncertainties vanish when the velocity of
test particle reach  the velocity of light, while the corresponding uncertainty
on time and length tends to infinity, but their product remains finite. we
have obtained the above uncertainty relations for massive test particles.
Note that the existence of minimum spatial  and causal structures also will
be appearer in seeking for quantum gravity such as the loop quantum gravity
\cite{loop} or string theory \cite{string}. The modification of special relativity
in which a minimum length, which may be the Planck length, joins the speed of light as an invariant is done in Ref. \cite{smolin}. 
We now discuss about light quanta or massless particles. In this case we have $u_\mu u^\mu=0$ and therefor equation (\ref{1-26}) becomes
\begin{eqnarray}\label{1-42}
\frac{D^2\xi}{D\tau^2}=-k^2e^{-2\pi kr}u_\gamma u^\mu\xi^\gamma.
\end{eqnarray}
If we assume a solution like $\xi^\mu= f^\mu(\tau)$, then by inserting 
into the above equation and by considering null conditionality for 4-velocity
we obtain $d^2f^\mu/d\tau^2=0$ and consequently
\begin{eqnarray}\label{1-43}
\xi^\mu=A^\mu\tau+B^\mu,
\end{eqnarray}
where $A^\mu$ and $B^\mu$ are constants of integration. This result shows that the extension of the massive test particle case to the photons
is not correct.
The above solution shows classically propagating massless particles in parallel
or cross propagating photons.
Note that the case of massless particles can be drive in this approach and
the Wesson suggestions \cite{4} can not lead us to this result. In fact
difference behavior of photons are proceed from confinement of gauge fields on the brane. Also 
As we know, the concepts of time in general relativity and quantum theory differ  intensely from each other. Time in quantum theory is an external
parameter, whereas  in general relativity time is dynamical one.
Consequently, a consistent theory of quantum gravity should exhibit a new concept of time. In general relativity spacetime is dynamical and therefore there is no absolute time. Spacetime influences material clocks in order to allow them to show proper time. The clocks, in turn, react on the metric and change the geometry \cite{Zeh}. In this sense, the metric itself is a clock. A quantization of the metric can thus be interpreted as a quantization of the concept of time. In this paper we showed that the consistency of geodesic
and geodesic deviation equations on the RS brane dictates the quantization
of proper time or clock rate. Note that this quantity cannot be dealt with as operators in ordinary quantum theories. The advantage of
this model is that it makes General Relativity compatible with de Broglie ideas, allows a geometric interpretation of de Broglie waves without any generalization of Riemannian spacetime. In this direction the problem needs
more survey.


\end{document}